\documentstyle[aps,prl,amsfonts]{revtex}
\begin{document}
\draft

\newcommand{\be}{\begin{equation}}
\newcommand{\ee}{\end{equation}}
\newcommand{\de}{\partial}
\newcommand{\too}{\longrightarrow}
\newcommand{\dyy}{\displaystyle}
\newcommand{\lps}{\langle}
\newcommand{\rps}{\rangle}
\newcommand{\dd}{\quad}
\newcommand{\bena}{\begin{eqnarray}}
\newcommand{\eena}{\end{eqnarray}}
\newcommand{\lpq}{\left[}
\newcommand{\rpq}{\right]}
\newcommand{\lpt}{\left(}
\newcommand{\rpt}{\right)}

\draft
\twocolumn
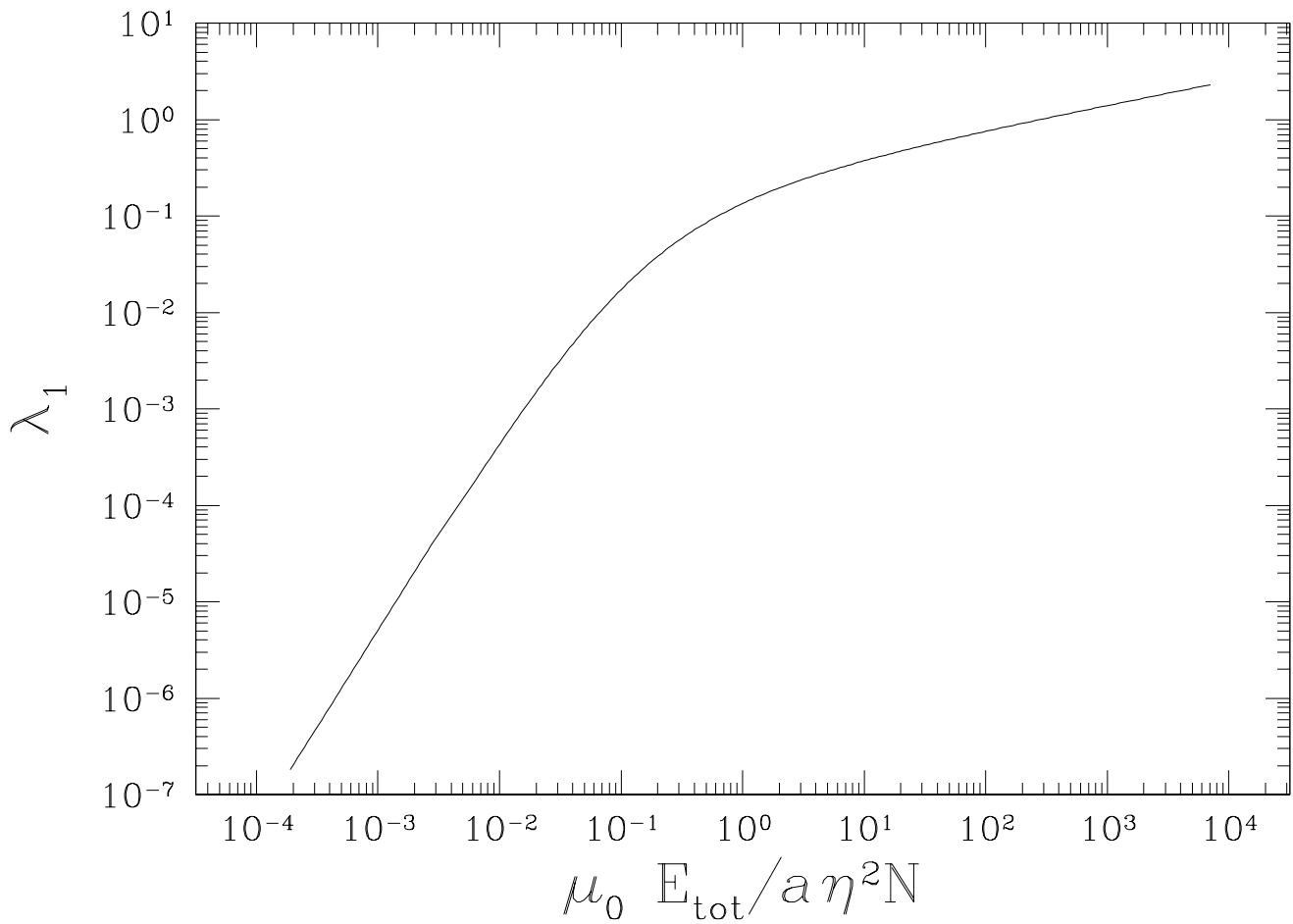
\title{\bf Analytic Lyapunov exponents in a classical nonlinear field
           equation}

\author{Roberto Franzosi$^{1,}$\cite{rob}, Raoul Gatto$^{2,}$\cite{raoul},
Giulio Pettini$^{1,}$\cite{giul}, and Marco Pettini$^{3,}$\cite{marco}\\}
\address{$^1$Dipartimento di Fisica dell'Universit\`a, Largo E. Fermi 2,
50125 Firenze, Italy\\
$^2$Departement de Physique Th\'eorique, Universit\'e de Geneve, 24 Quai
Ernest-Ansermet, CH-1211 Geneve, Switzerland\\
$^3$Osservatorio Astrofisico di Arcetri, Largo E. Fermi 5,
50125 Firenze, Italy}
\date{\today}
\maketitle

\begin{abstract}
It is shown that the nonlinear wave equation $\partial_t^2\varphi -
\partial^2_x \varphi -\mu_0\partial_x(\partial_x\varphi)^3 =0$, which is the
continuum limit of the Fermi-Pasta-Ulam (FPU) $\beta$ model, has a positive
Lyapunov exponent $\lambda_1$, whose analytic energy dependence is given.
The result (a first example for field equations) is achieved by evaluating 
the lattice-spacing dependence of $\lambda_1$ for the FPU model  within the 
framework of a Riemannian description of Hamiltonian chaos. 
We also discuss a difficulty of the statistical mechanical treatment of this
classical field system, which is absent in the dynamical description.
\end{abstract}
\pacs{PACS: 05.45.+b; 03.40.-t; 03.40.Kf}
The numerical study that E.Fermi, J.Pasta and S.Ulam made at Los Alamos,
on a chain of
anharmonically coupled oscillators, represents a milestone in the development
of the modern theory of nonlinear dynamics \cite{FPU}.
The surprising outcomes of this
first computer experiment in the history of physics have been very seminal.
Among the other attempts to explain the apparent lack of thermalization of
the energy initially stored in one of the linear modes of the chain,
Zabusky and Kruskal \cite{Zabusky,Zabusky1} remarked that a continuum
limit version of the
Fermi-Pasta-Ulam (FPU) model
 leads to the Korteweg-de-Vries (KdV) equation, when the
$\alpha$ model is considered, and to a modified KdV equation when the
$\beta$ model is considered. These are nonlinear, integrable partial
differential
equations where a
special class of solitary waves -- that Zabusky and Kruskal called {\it
solitons} -- exists and can to some extent explain the FPU recurrences
\cite{Tuck}.

In the present work we tackle the FPU-$\beta$ model, described by the
Hamiltonian
\begin{equation}
H(p,q) = \sum_{i= 1}^N\left[ \frac{1}{2} p_i^2 +
\frac{1}{2} (q_{i+1} - q_i)^2 + \frac{\mu}{4}
(q_{i+1} - q_i)^4 \right],
\label{fpu}
\end{equation}
and we consider another legitimate continuum limit \cite{cercignani}
of this system, described by the Hamiltonian
\begin{equation}
H(\pi,\phi) = \int_0^L\,d x \left[\frac{1}{2\nu} \pi (x)^2 +
\frac{\eta}{2} (\nabla\phi)^2 + \frac{\mu_0}{4}
(\nabla\phi)^4 \right]~,
\label{fpu-cont}
\end{equation}
and leading to the field equation (setting $\eta =\nu =1$)
\begin{equation}
\frac{\partial^2\phi}{\partial x^2} - \frac{\partial^2\phi}{\partial t^2} +
\mu_0\frac{\partial}{\partial x}\left(\frac{\partial\phi}{\partial x}
\right)^3=0~.
\label{eq-cont}
\end{equation}
Also this continuum limit of the FPU model might appear integrable on the
 basis of the following reasoning \cite{weiss}.
Let us consider the Legendre transform:
$\chi =\partial_x\phi$, $\tau =\partial_t\phi$ and $x =\partial_\chi\psi $,
$t=\partial_\tau \psi$ with the relation $\psi + \phi =x\chi +t\tau$, whence
$\partial_x^2\phi =D\partial_\tau^2\psi$, $\partial_t^2\phi =
D\partial_\chi^2\psi$, where $1/D=\partial_\chi^2\psi\partial_\tau^2\psi -
(\partial_{\chi\tau}^2\psi )^2$. Substituting into Eq.(\ref{eq-cont}) we can
linearize it to
\begin{equation}
\frac{\partial^2\psi}{\partial\chi^2} = (1 +3 \mu_0\chi^2)
\frac{\partial^2\psi}{\partial \tau^2}
\label{eq-lin}
\end{equation}
which can be solved by variable separation by setting $\psi =F(\chi)G(\tau)$,
where $G^{\prime\prime}\pm c^2G=0$, and $F^{\prime\prime}\pm
c^2(1+3\mu_0\chi^2)F=0$.
In principle, by inverting this Legendre transform, Eq.(\ref{eq-cont})
could be analytically solved. However, as we shall discuss,
the method sketched above can only lead at most to {\it local} and not to
{\it global} invertibility, both in time and space. Actually we are going to
show that the field equation (\ref{eq-cont}) for the FPU $\beta$ model
in the continuum limit is chaotic (the Lyapunov characteristic exponent of
the solutions is positive).
The pattern of the Lyapunov exponent that we shall derive is shown in
Fig.1.
The cross-over in the energy dependence of the largest Lyapunov exponent has
been attributed in Refs.\cite{PettiniLandolfi} to a (smooth)
transition between weak and strong chaos \cite{KAM}. 
The persistence of such a cross-over
 also in the continuum limit is a remarkable fact. Loosely speaking, at high
energy, in the strongly chaotic regime, the spatio-temporal behavior of the
field $\phi (x,t)$ should look like a fully developed ``turbulent'' field,
whereas this should not be the case at low energy, where a weakly
``turbulent'' dynamics would set in.
Our calculation shows that, no matter whether weak or strong, chaos is
present in the system at any energy, and since the system is Hamiltonian, 
we can believe that a standard statistical mechanical description is allowed.
This raises an important problem that we now discuss.

For the lattice system
corresponding to
(\ref{fpu-cont}) the standard canonical partition function
$Z=\int \prod_{i=1}^N d\pi_i d\phi_i \exp \left[ -\beta H(\{\pi_i\},
\{\phi_i\}) \right]$
in the limit of lattice spacing $a\rightarrow 0$ becomes
\begin{equation}
{\cal Z}=\int {\cal D}\pi (x)\,{\cal D} \phi (x)e^{ -\beta \int_0^L
 dx \left[ \frac{1}{2}\pi^2 + \frac{1}{2}(\nabla\phi )^2+\frac{\mu_0}{4}
(\nabla\phi)^4\right]}~~.
\end{equation}
It is interesting at this point to raise a problem that comes out when some
expected dynamical properties of the field $\phi (x,t)$ are compared to their
statistical counterparts worked out by means of ${\cal Z}$.
By means of an orthogonal change of coordinates, the Hamiltonian (\ref{fpu})
is rewritten in the form \cite{PettiniLandolfi}
$H(P_k,Q_k)=\sum_k E_k=\sum_k [\frac{1}{2}(P_k^2+\omega_k^2Q_k^2)+
\mu\sum_{k_1,k_2,k_3}C(k,k_1,k_2,k_3)Q_kQ_{k_1}Q_{k_2}Q_{k_3}]$ where the
summations run from $k=1$ to $k=k_{max}$ for a lattice, and from $k_0$ to
$k_{max}=\infty$ in the continuum limit. In the new variables the partition
function becomes ${\tilde Z}=\int \prod_{k=1}^{k_{max}} dP_k dQ_k \exp
\left[ -\beta H(\{P_k\},\{Q_k\}) \right]$ which, in the limit $k_{max}
\rightarrow\infty$, gives ${\tilde{\cal Z}}=\int {\cal D}P(k){\cal D}Q(k)
\exp \left[ -\beta\int_{k_0}^\infty dk H(P(k),Q(k))\right]$, where
$k_0=\frac{2\pi}{L}>0$ for a finite support. The generalized equipartition
theorem \cite{munster}  states that $\langle Q_k\partial H/\partial Q_k\rangle
= const$ independently of $k$, and from Ref.\cite{Casartelli} we know that
for the FPU $\beta$ model such an average is  $\langle Q_k\partial H/\partial
Q_k\rangle \simeq k^2Q_k^2(1+\alpha)$, where $\alpha$ is a constant,
and therefore
$Q_k^2\simeq k^{-2}(1+\alpha )^{-1}$, which, being independent of $k_{max}$,
still holds true in the continuum limit, i.e
$\vert\hat\phi(k)\vert^2\simeq k^{-2}(1+\alpha )^{-1}$.
Whence an ultraviolet catastrophe,
i.e. the divergence of average total energy when $k_{max}\rightarrow\infty$
also on a finite support $[0,L]$. However such a difficulty is absent in the
dynamical equation (\ref{eq-cont}), which, being derived from a Hamiltonian,
must keep the energy constant, therefore finite if it is finite at time $t=0$.

The point is that statistical mechanics and dynamics yield very different
{\it large}-$k$ spectral properties. In fact, already on the basis of the
purely dimensional analysis discussed in
Refs.\cite{lprv}, we can easily find that in order to bound the total energy
(on a finite support $[0,L]$) to a finite value, the ultraviolet asymptotic
power spectrum of $\phi (x)$ must be bounded above by $\vert\hat\phi (k)
\vert^2\sim k^{-3}$, which means that the high spatial-frequency modes 
cannot have the same energy of the low spatial-frequency modes.

The rough estimate of the
ultraviolet spectrum of the continuum model can be improved as follows. Any
smooth initial condition $\phi (x,0)\in{\cal C}^\infty ({\Bbb R})$ of the
equation of motion (\ref{eq-cont}) has to remain smooth at any further
time $t$, i.e. $\phi (x,t)\in{\cal C}^\infty ({\Bbb R})$. Now, the analytic
continuation $\Phi (z,t)$ of $\phi (x,t)$ will be of class ${\cal C}^\omega$
on a strip whose width $\delta$ will be determined at any given time $t$ by
the distance of the closest singularity to the real axis. If
$z^s_n=x^s_n + i y^s_n$  are the locations of the singularities of
$\Phi (z,t)$, to the leading order one finds \cite{Carrier,Frisch}
$\hat\Phi (k)\sim k^{-\alpha}\sum_n c_n e^{ikz^s_n}$  for
$k\rightarrow\infty$, where, apart from the power-law prefactor $k^{-\alpha}$,
the sum is carried over the $Im\,z >0$ half plane and $c_n$ are suitable
coefficients. The factors $e^{ikz^s_n}=e^{ikx^s_n}e^{-ky^s_n}$ efficiently
single out the closest singularity to the real axis to give $\vert\hat\Phi
(k)\vert^2\le A e^{-\delta k}$ at $k\rightarrow\infty$ which provides a better
ultraviolet estimate of $\vert\hat\phi (k)\vert^2$ stemming from the 
constraint of differentiability of $\phi (x)$. Needless to say, this 
ultraviolet exponential fall off of $\vert\hat\phi (k)\vert^2$ guarantees 
the finiteness of the total energy for any finite support $[0,L]$.
It is not unreasonable to think that a relationship might exist between the
minimum width of the spatio-temporal cell where the Legendre transform of
Eq.(\ref{eq-cont}) - sketched at the beginning of this paper - is invertible,
 that is where Eq.(\ref{eq-cont}) is locally integrable, and the width of the
$(k,\omega (k))$ cell where the power spectrum $\vert\hat\phi (k)\vert^2$
exponentially falls off. There is apparently no way of reproducing the
ultraviolet exponential decay of $\vert\hat\phi (k)\vert^2$
-  which is naturally brought about by the dynamical equation
(\ref{eq-cont}) - within the statistical mechanical framework. The field
$\phi (x,t)$ can be chaotic in space and time only down to some small scale
which naturally provides a cutoff that, inserted into $\tilde{\cal Z}$, would
remove the mentioned divergences.
Let us mention that the physical meaning of a rigorous treatment
of another classical nonlinear field equation (complex Ginzburg-Landau)
\cite{eck1} is coherent with the theoretical scenario depicted above.

To come now to our main result, chaoticity in the continuum limit, we
calculate the Lyapunov exponent by extending to such a limit a method to
tackle Hamiltonian chaos that has already given
excellent analytic predictions of the largest Lyapounov
exponent in the thermodynamic limit of the FPU $\beta$-model \cite{CCP} on
a lattice. This method exploits the mathematical identification of an
Hamiltonian flow with the geodesic flow on a suitable Riemannian 
``mechanical manifold'' consisting of an enlarged configuration spacetime
($\{q^0\equiv t,q^1,\ldots,q^N\}$
plus one real coordinate $q^{N+1}$), whose arc-length 
$ds^2 =  -2V({\bf q}) (dq^0)^2 + a_{ij} dq^i dq^j
+ 2 dq^0 dq^{N+1}~$ defines the so-called Eisenhart metric 
$g_{_E}$ \cite{Eisenhart}; $V$ is the potential.
In the geometrical framework, the (in)stability
of the trajectories is the (in)stability
of the geodesics, and it is completely determined by the
curvature properties of the underlying manifold according to
the Jacobi equation \cite{doCarmo} for the geodesic deviation. This
equation, written for Eisenhart metric, entails the 
usual tangent dynamics equation $\ddot\xi^i+(\partial^2V/\partial q_i
\partial q^j)\xi^j=0$, which is used to
measure Lyapunov exponents in standard Hamiltonian systems.
Having recognized its geometric origin, it can be transformed, under
certain hypotheses \cite{CCP}, into 
an {\it effective} scalar stability equation that, {\it independently} of the
knowledge of dynamical trajectories, provides a  measure of their
average degree of instability. 
This effective stability equation is in the
form of a stochastic oscillator equation
$\,{\ddot\psi} + [\kappa_0 +\sigma_\kappa\eta (t)] \, \psi = 0~$ \cite{CCP},
where $\kappa_0 =  {\langle K_R \rangle}/{N}$,
$\sigma^2_\kappa  =  {\langle (K_R - \langle K_R \rangle)^2 \rangle}/{N}~$,
and $K_R\equiv \Delta V = \sum_{i=1}^N \partial^2 V/\partial q_i^2$ is the
Ricci curvature of the mechanical manifold, computed with $g_E$;
$\eta (t)$ is a gaussian $\delta$-correlated random process of unit variance.
The exponential
growth rate $\lambda$ of $(\dot\psi^2+\psi^2)$ is 
computed exactly to provide the following estimate of the largest Lyapunov
exponent:
\begin{equation}
\lambda = \frac{\Lambda}{2} - \frac{2 \kappa_0}{3 \Lambda}\,,~~~~
\Lambda = \left[2\sigma_\kappa^2 \tau +
\sqrt{{\frac{4 \kappa_0}{3}}^3 + 4\sigma_\kappa^4 \tau^2}~\right]^\frac{1}{3}
\label{lambda}
\end{equation}
where
$\tau = \pi\sqrt{\kappa_0}/(2[\kappa_0(\kappa_0 + \sigma_\kappa)]^
{\frac{1}{2}} +\pi\sigma_\kappa )$
is a characteristic time scale worked out on the basis of geometrical
arguments.
In the limit $\sigma_\kappa/\kappa_0 \ll 1$ one finds
$\lambda \propto \sigma_\kappa^2$ \cite{CCP}.

In this geometric picture chaos is mainly originated by the parametric
instability activated by the fluctuating curvature ``felt''
by the geodesics.

The quantities $\kappa_0$ and $\sigma^2_\kappa$ can be exactly computed for
the FPU-$\beta$ model as microcanonical averages by taking advantage of the
analytically known canonical partition function and by using the conversion
formulas relating canonical and microcanonical averages. The final analytic
expressions are given in Ref.\cite{CCP}, and among them we report those 
needed here
\begin{eqnarray}
	 \frac{1}{N} \lps K_R  \rps^\mu  (\theta)& =& 2 + \dyy\frac{3}{\theta}
	 \dyy\Delta(\theta ) \\ \label{Omega0}
        \dyy\frac{1}{N} \lps\delta^2 K_R \rps^\mu  (\theta)& = &
	 \dyy\frac{1}{N}  \dyy\frac{9}{\theta^2} \left[
	2 - 2 \theta \cdot \dyy\Delta (\theta) -
	 \dyy\Delta^2(\theta) \right]\nonumber\\
          & - &
	 \dyy\frac{\beta^2}{c_v(\theta)}
	 \lpt \dyy\frac{\de \lps K_R/N\rps^G}{\de \beta}
	 \rpt^2\\ \label{fluttK}
	 \epsilon (\theta)& =& \dyy\frac{1}{8 \mu} \lpq \dyy\frac{3}{\theta^2}
        + \dyy\frac{1}{\theta}\Delta(\theta ) \dyy
		 \rpq   \dd  ,
\label{edens}
\end{eqnarray}
where $\Delta(\theta )=\frac{D_{-3/2}(\theta)}{D_{-1/2}(\theta)}$, with
$D_{-3/2}$ and $D_{-1/2}$ parabolic cylinder functions, the
parameter $\theta$ is a function of the inverse temperature $\beta$ through
$\theta = (\beta /2 \mu)^{1/2}$, and  $\epsilon (\theta)$ is the energy per
degree of freedom of the system.
The microcanonical average of curvature fluctuations involves a correction
term to their canonical average which involves the specific heat $c_v(\theta)$
computed, as usual, as the second derivative of the free energy.
These quantities enter the formula (\ref{lambda}) to give the analytic values
 of Lyapunov exponents at different energies.
Now, it is useful to work out simple analytic expressions
of $\lambda_1(\epsilon)$ at low and high energy densities. In order to obtain
them, notice that from Eq.(\ref{edens}) one gets $\mu\epsilon =f(\theta)$,
a function of $\theta$; this function is invertible, because in the absence of
phase transitions $\epsilon (T)$ ( $T$ is the temperature ) is always
one-to-one. Therefore we can write $\theta =f^{-1}(\mu\epsilon )$ and
consequently $\kappa_0$ and $\sigma^2_\kappa$ can be expressed as functions of
$\mu\epsilon$. Making use of the asymptotic expressions of the parabolic
cylinder functions, this inversion is easily obtained in the two opposite
limits
$\epsilon\rightarrow 0$ and $\epsilon\rightarrow\infty$. Also simple analytic
expressions for $\kappa_0$, $\sigma^2_\kappa$ and $\tau$ can be obtained in
these two limits. At $\epsilon\rightarrow 0$, and $\theta\gg 1$, we get
\begin{eqnarray}
\kappa_0&=& 2 + 6\mu\epsilon - \frac{27}{2}\mu^2\epsilon^2 +
           {\cal O}(\epsilon^3)\\
\sigma^2_\kappa &=& 36\mu^2\epsilon^2 + {\cal O}(\epsilon^3)~~.
\end{eqnarray}
Let us remark that there is
some arbitrariness in the derivation of $\tau$ in Ref.\cite{CCP}; however from
Eq.(\ref{lambda}) we see that the only relevant constraint on $\tau$ is that
$\sigma^2_\kappa\tau\rightarrow 0$ for $\epsilon\rightarrow 0$, so that
$\Lambda$ -- and $\lambda_1$ with it -- can be expanded in series of the small
quantity $\sigma^2_\kappa\tau$ yielding
\begin{equation}
\lambda_1(\epsilon )=\frac{\sigma^2_\kappa\tau}{2\kappa_0} + \dots= 9\mu^2
\epsilon^2 \tau + \dots
\end{equation}
which, with the explicit computation of $\tau$ gives
\begin{equation}
\lambda_1(\epsilon)=\frac{9}{4}\frac{\pi}{\sqrt{2}}\mu^2\epsilon^2 + {\cal O}
(\epsilon^3)
\end{equation}
which is in strikingly good agreement with the low energy ``experimental''
results for $\lambda_1$.
Similar developments can be worked out in the limit $\epsilon\rightarrow
\infty$ (and so $\theta\ll 1$) giving
\begin{equation}
\kappa_0\simeq \frac{4\pi}{\Gamma^2\left(\frac{1}{4}\right)}
            \sqrt{24\mu\epsilon}\ ,~~~~ 
\sigma^2_\kappa\simeq 16\mu\epsilon \left( 3-\frac{32\pi^2}{\Gamma^4
\left(\frac{1}{4}\right)}\right)
\end{equation}
by computing again $\tau(\epsilon)$  we find
$\tau (\epsilon )\simeq c(\mu\epsilon)^{-1/4}$ yielding
\begin{equation}
\lambda_1(\epsilon )\simeq c^\prime(\mu\epsilon)^{1/4}
\end{equation}
where $c$ and $c^\prime$ are constants. Also this scaling law is in perfect
agreement with numerical results.
The comparison between the full analytic prediction
of $\lambda_1(\epsilon )$, obtained by substituting Eqs.(\ref{Omega0}) and
(\ref{fluttK}) into
Eqs.(\ref{lambda}), shows a perfect agreement
 with the outcomes of a standard numerical
computation of the largest Lyapunov exponent at different values of 
$\epsilon$.
Let us remark that the analytic prediction of $\lambda_1(\epsilon)$ is 
obtained
in the $N\rightarrow\infty$ limit, which is implicit in the computation of the
microcanonical averages of the Ricci curvature and of its fluctuations.
Now, as the analytic expression in (\ref{lambda}) is in excellent agreement
with the numerical data obtained
for discrete (particle) systems at different values of $N$, there is no
reason to doubt that such an agreement will hold true also when $N$ is varied
while keeping the total length $L$ of the system fixed, i.e. in the
{\it continuum limit}. The non-trivial difference between this limit and
thermodynamic limit in Ref.\cite{CCP} is that $N\rightarrow\infty$ while the
energy remains finite in the former case, whereas both $N$ and energy 
diverge in the latter case.  Thus
let us now consider the Hamiltonian (\ref{fpu-cont}) discretized on a lattice
of length $L$ and spacing $a$ (with $L=Na$)
\begin{eqnarray}
H(\{\pi_i\},\{\phi_i\}) & = & \nonumber \\
\sum_{i= 1}^N a\left[ \frac{\pi_i^2}{2\nu}\right. & + &\left.
\frac{\eta}{2}\left(\frac{\phi_{i+1} - \phi_i}{a}\right)^2 + \frac{\mu_0}{4}
\left(\frac{\phi_{i+1} - \phi_i}{a}\right)^4\right]~,
\label{fpu-a}
\end{eqnarray}
where the dimensional constants $\nu,\eta$ have been introduced.
We have thus obtained a discrete (lattice) system, similar to that described
by the Hamiltonian (\ref{fpu}); in order to make a direct comparison between
 the
two Hamiltonians of Eqs.(\ref{fpu-a}) and (\ref{fpu}), we must now absorbe the
dimensional parameter $a$ into the constants of the Hamiltonian. By means
of the following rescaling of the coordinates $\phi_i$ and of the time $t$:
$\xi_i=\phi_i/\alpha$, with $\alpha=\sqrt{a/\eta}$, $t^\prime=t/\gamma$ and
$\gamma =\sqrt{\eta/\nu}$, and denoting by $\{\zeta_i\}$ the momenta 
conjugated to $\{\phi_i\}$, $H$ of Eq.(\ref{fpu-a}) is put in the form
\begin{equation}
H_a(\{\zeta_i\},\{\xi_i\}) = \sum_{i= 1}^N \left[ \frac{\zeta_i^2}{2} +
\frac{1}{2}(\xi_{i+1} - \xi_i)^2 + \frac{\mu_0}{a\eta^2}
\frac{1}{4}(\xi_{i+1} - \xi_i)^4\right]~,
\label{fpu-b}
\end{equation}
which is just the form of $H$ in Eq.(\ref{fpu}), so that we can apply the
geometric treatment of dynamical chaos also to the systems described by
this family of Hamiltonians $H_a$. The precise meaning of this comparison
is the following. The continuum Hamiltonian (\ref{fpu-cont}) has $H$ of Eq.
(\ref{fpu-a})
as its discretized version, where the parameters $\eta,\, \nu,\, \mu_0$
are equal to the values of the continuum model. After the recasting of
Eq.(\ref{fpu-a}) into the form (\ref{fpu-b}) we can easily derive the 
value of the coupling constant $\mu$ of the discrete FPU model (\ref{fpu})
so that a correspondence can be made between the discrete model and the
lattice version (\ref{fpu-b}) of the continuum model for any value of the
lattice spacing $a$. Evidently the relation between the coupling constants is
$\mu=\mu_0/a\eta^2$,
and this tells that in order to represent the solutions of a set of systems
(\ref{fpu-a}) by means of those of a set of discrete systems (\ref{fpu})
with a finer and finer sampling of the spatial support of length $L$,
i.e. letting $a\rightarrow 0$, the coupling constant $\mu$ must be larger
and larger.
To the lattice-discretized system (\ref{fpu-b}) we can now apply
all the above  equations for the geometric description of chaos,
provided that we replace $\mu$ with $\mu_0/a\eta^2$.
In passing to the continuum limit we have to let $a\rightarrow 0$ while the
total energy of the system $E_{tot}$ is kept fixed.
Hence the energy density $\epsilon$ has to diminish as $E_{tot}/N$, but
$N=L/a$ and then $\epsilon = E_{tot} \cdot a /L$. Hence we obtain
        \be
	\mu \cdot \epsilon =\frac{\mu_0 E_{tot} a}{a \eta^2 L} = cost
	\dd .
	\label{mue}
	\ee
From Eq.(\ref{mue}) it then follows that the equations for $\kappa_0$ and
$\sigma^2_\Omega$ (\ref{Omega0})-(\ref{edens}) are stable with respect to
the limit to the continuum and so is the largest Lyapunov exponent.
Therefore {\it the field equation (\ref{eq-cont}) is chaotic} and its Lyapunov
exponent has the pattern shown by Fig.1. This constitutes a first example of
analytic calculation of $\lambda_1(\epsilon )$ for a field equation.
Implicit in our computation is the possibility of approximating, as accurately
as needed, a partial differential equation by means of a finite (truncated)
system of ordinary differential equations, for which Lyapunov exponents are 
well defined.
In conclusion we have shown that the
continuum limit of the Fermi-Pasta-Ulam $\beta$ model is chaotic. We have
discussed the interesting problematics raised by this result on the
statistical description of classical field theory models that might be 
relevant also to the Wick-rotated quantum field theories.

It is a pleasure to thank M. Casartelli for useful comments.

\begin{figure}
\caption{Analytic Lyapunov exponent for the continuum limit as a function of 
the physical parameters. }
\label{fig1}
\end{figure}


\begin{references}
\bibitem{weiss} We are indebted to J.Weiss for having drawn our attention
                to this method.
\bibitem{KAM} Note that, at arbitrarily large $N$, no sizeable region of 
regular motions is left in phase space. KAM theorem predicts the existence 
of such regions only below exceedingly tiny amplitudes of the integrability 
breaking perturbation, moreover vanishing exponentially fast with $N$.
\end{references}
\end{document}